\documentstyle[prl,aps,epsfig]{revtex}

\begin{document}

\twocolumn[\hsize\textwidth\columnwidth\hsize\csname
@twocolumnfalse\endcsname

\title{Vortex Softening: Origin of the second peak effect in Bi$_2$Sr$_2$CaCu$_2$O$_{8+\delta }$}

\author{V.F. Correa, G. Nieva and F. de la Cruz}

\address{Comisi\'on Nacional de Energ\'{\i}a At\'omica,
Centro At\'omico Bariloche, 8400 S. C. de Bariloche,
Argentina}

\date{\today}

\maketitle

\begin{abstract}

Transverse ac permeability measurements in Bi$_2$Sr$_2$CaCu$_2$O$_{8+\delta }$ single crystals 
at low fields and temperatures in a vortex configuration free of external forces show that the 
decrease of the critical current as measured by magnetization loops  
at the second peak effect is an artifact due to creep. On the other hand, the increase of 
critical current at the second peak is due to a genuine softening of the tilting elastic 
properties of vortices in the individual pinning regime that precedes the transition to a 
disorder state.

\end{abstract}

\pacs{74.60.Ec, 74.60.Ge, 74.60.Jg, 74.72.Hs}
 
\vskip1pc] \narrowtext

The existence of the second peak in the low field low temperature  
magnetization of Bi$_2$Sr$_2$CaCu$_2$O$_{8+\delta }$ (Bi2212) as
well as the peak effect observed in the critical current J$_c$ of low 
(LTS) and high temperature superconductors are manifestations of 
instabilities of the vortex structure (VS) in the presence of pinning
potentials. Since the first explanation by Pippard and the theoretical
description by Larkin and Ovchinnikov \cite{Larkin} several other
alternatives have been proposed, including order disorder thermodynamic
phase transitions \cite{LeDoussal} \cite{Daemen} \cite{Zeldov} 
\cite{Gaifullin} \cite{Horovitz}.

The most remarkable manifestation in both phenomena is that J$_c$
goes through a maximum when increasing field. Thus, most
experiments studying the anomaly are based in measurements of 
J$_c$ and consequently describe the properties of a non
equilibrium thermodynamic state. In particular, in the case of the
second peak in Bi2212, J$_c$ determined by magnetization
loop measurements is strongly affected by time dependent phenomena.
As a result, it has been suggested \cite{Niderost} a possible explanation
of the second peak effect as due to different relaxation rates of the vortex 
system in a non homogeneous field distribution induced by the critical 
state. Other suggestion supporting the dynamical origin of the 
phenomenon has been introduced \cite{Baziljevich} from local magnetization
loops induced in short time scales. This is in contrast with other 
experiments in Bi2212 where a thermodynamic phase transition is claimed 
\cite{LeDoussal} \cite{Daemen} \cite{Zeldov} \cite{Gaifullin} \cite{Horovitz} 
to be associated with the second peak.

In this paper we compare experimental results obtained by different 
techniques in order to distinguish the possible dynamical contribution to 
the second peak from that caused by genuine changes in the elastic 
response of the VS in the presence of pinning potentials.
We have compared magnetization loop measurements in the critical state
with results obtained from ac transverse permeability of vortex configurations
free of bulk magnetic gradients. 
With this last constrain we have been 
able to detected an enhancement of the intrinsic pinning potential of the 
vortex lattice in a region of fields where the second peak is detected.
The observed behavior is due to a softening of the elastic properties of 
vortices that might be considered a precursor of a phase transition.

The Bi2212 single crystals used in this work were grown using the 
self flux technique \cite{Kaul} and they have typical dimensions 
2x1x0.02 mm$^3$.  
We have made measurements of the field cooled, FC, ac transverse 
permeability in the Campbell limit, where the VS remains
pinned under the perturbation induced by the ac field. A sketch of 
the experimental configuration with the applied field H$_a$ in the 
$c$ direction can be seen in the insert of Fig. 1 and details in 
ref. \cite{Arribere}. 

The ac permeability in this configuration is given \cite{Brandt} by

\begin{equation}
\mu \, = \, \frac { 2 \, \lambda_{ac}}{d} \,  tanh(\frac {d}{2\, \lambda_{ac}})
\end{equation}

\noindent where d is the thickness of the sample and the ac penetration depth 
$\lambda_{ac}$ in the Campbell limit follows expression \cite{Brandt}

\begin{equation}
\lambda_{ac}^{2} = \lambda_{ac}^{2}(H_a=0) + \lambda_{C}^{2} 
\end{equation}

\noindent where $\lambda_{C}$ is the Campbel penetration depth.
The Campbell limit is achieved when the ac response is characterized 
\cite{Campbell} by vortices locked in a pinning potential, linear frequency
independent response to ac excitations and a very small dissipation due to
the displacement of the vortex core within the effective pinning potential.
The Campbell penetration depth is given by \cite{Brandt}  

\begin{equation}
\lambda_{C}^{2}(H_a,T) = \frac{c_{44}}{\alpha_L (H_a,T)} 
\end{equation}

\noindent The elastic constant c$_{44}$ determines the vortex response to the 
tilting force induced in the transverse configuration. The Labusch 
parameter $\alpha_L$ (H$_a$,T) is the effective pinning potential responding
to the ac excitation.

It has been demostrated \cite{Blatter} that the pinning barrier is a 
function of the force acting on the vortices.
When the only external force applied to the VS is that
of the perturbation induced by the ac field the elastic response in
the Campbell limit is determined by the curvature at the bottom of the
effective pinning potential. This requires the electromagnetic force induced
by the ac field to be much smaller than the critical one to remove
vortices from the pinning sites. 
Basically, the VS localized at the bottom of the Labusch
potential represents a free force VS in thermodynamic
equilibrium with the pinning potential. In practice, the true equilibrium
state for a given external field and temperature can not be reached.
On the other hand, the structure obtained by freezing the FC vortex
system through the liquid solid-first order transition is essentially
a vortex free force configuration down to temperatures where the individual
pinning vortex limit is achieved \cite{Deco}. In this configuration 
the pinning barrier is maximum \cite{Blatter} and in 
the individual vortex pinning limit $\alpha_L$(H$_a$,T) becomes field independent. 
In this limit, the field  dependence of 
$\lambda_{C}^{2}$=  (H$_a$ $\phi_{0}$ / 4 $\pi$ $\alpha_L$(H$_a$,T)), is 
given by the field dependence of c$_{44}$ = H$_a$ $\phi_{0}$/4 $\pi$.
Typically, when the field is increased the vortex-vortex interaction
becomes relevant and a crossover to the collective pinning regime takes place.
Then $\alpha_L$(H$_a$,T) decreases, and the Campbell penetration depth increases.

Magnetization loops were also 
measured using a commercial Quantum Design SQUID magnetometer. From these 
measurements J$_c$ was extracted using the expression \cite{Bean}

\begin{equation}
J_c \, = \, \frac {3}{2} \, \frac {c \, \Delta M}{R} 
\end{equation}

\noindent where $\Delta$ M is the difference in magnetization for a given field 
and R is a typical dimension of the sample.
Equation (4) gives J$_c$ only when creep 
effects can be disregarded. In this case, J$_c \propto
\alpha_L$ (H$_a$,T). The measurements were taken in time scale of few 
minutes. 

In Fig. 1 we show the field dependence of $\lambda_{ac}^{2}$ at 15K,
normalized by the thickness of the sample. The imaginary part of
the permeability shows the low dissipation peak corresponding to the
Campbell limit, see the insert at the lower rigth of the figure.  The 
linear dependence of $\lambda_{ac}^{2}$ with H$_a$ confirms the individual
pinning limit up to fields as high as 1500 Oe.

\vspace{-15pt}

\begin{figure}[]
	\rotatebox{270}{\includegraphics[width=\columnwidth]{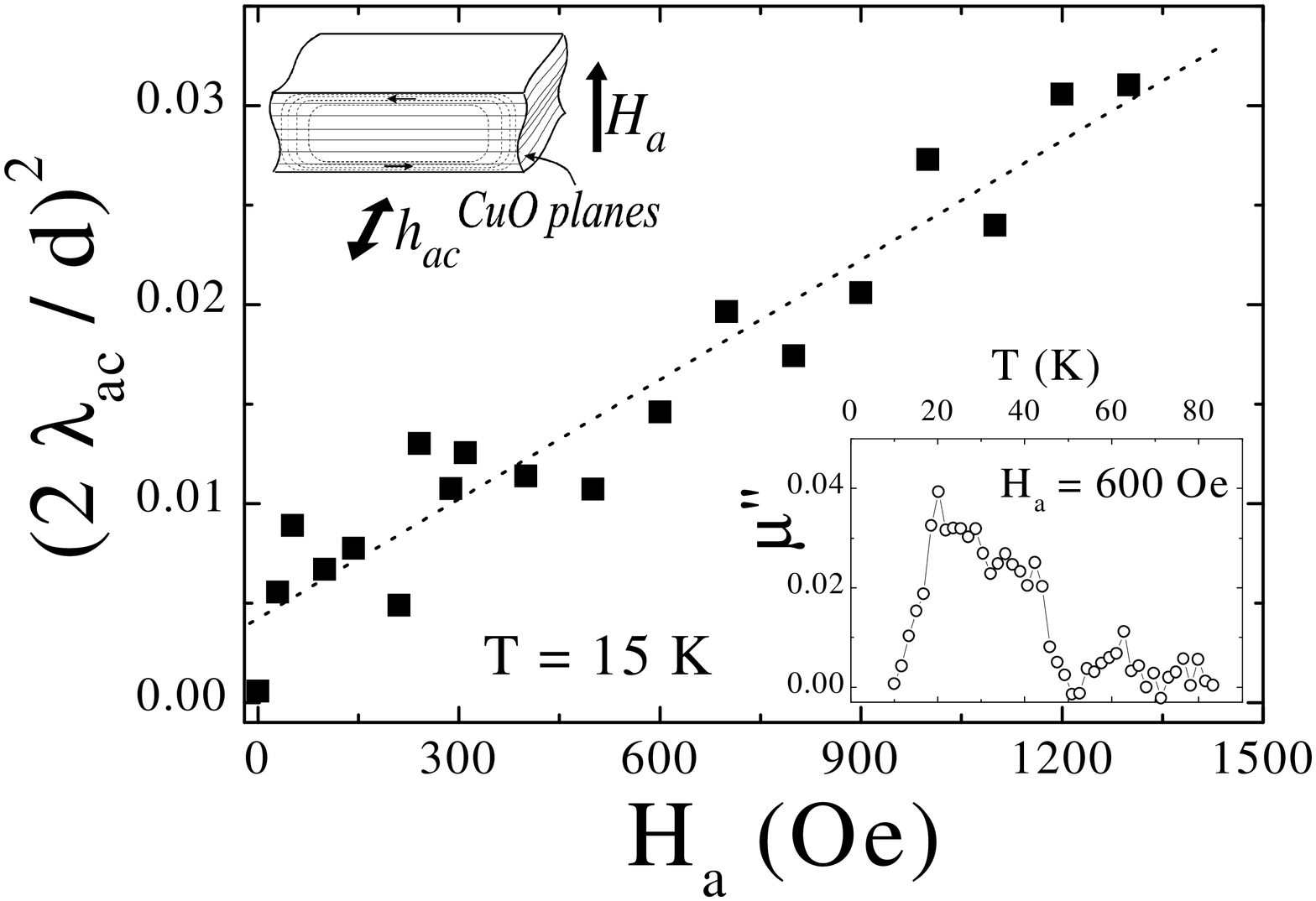}}
	\vspace{-60pt}
	\caption{Field dependence of $\lambda_{ac}^{2}$ at 15K,
    normalized by the thickness of the sample. The dotted line is a linear fit to the data. 
	Insert at the lower rigth: Imaginary part of
    the permeability  vs T at H$_a$ =600 Oe. Insert at the upper left: Experimental configuration.}
	\label{g1}
\end{figure}

The field independent Labusch coefficient can be extracted from the 
slope of the data or from each experimental value using the expression 

\begin{equation}
\alpha_L \, = \, \frac {H_a \phi_{0}}{4 \pi} \, (\lambda_{ac}^2 (H_a,T) - \lambda_{ac}^2 (0,T)) 
\end{equation}

\noindent In principle, $\lambda_{ac}$ (0,T) is the London penetration 
depth $\lambda_{L}$. However, it is often found \cite{pasquini} that 
$\lambda_{ac}$(0,T) obtained from the extrapolation of the linear field
dependence is larger than $\lambda_{L}$. In our case 
$\lambda_{ac}$(0,T) $\approx$ 3$\lambda_{L}$. This is due to a rapid increase of the
penetration depth in the low field range, H$_a$ $\leq$ 30 Oe,
of unknown origin, followed by the linear increase shown in Fig. 1. The rapid 
increase is incorporated in $\lambda_{ac}^{2}$ (0,T) as  a fitting parameter for 
the linear field dependence of $\lambda_{ac}^{2}$ (H$_a$,T). The slope of 
$\lambda_{ac}^{2}$ (H$_a$,T) determines $\alpha_L$ (H$_a$,T), plotted as a full
line in Fig. 2 together with the corresponding values 
obtained from expression (5). 

\begin{figure}[]
	\includegraphics[angle=90, width=\columnwidth]{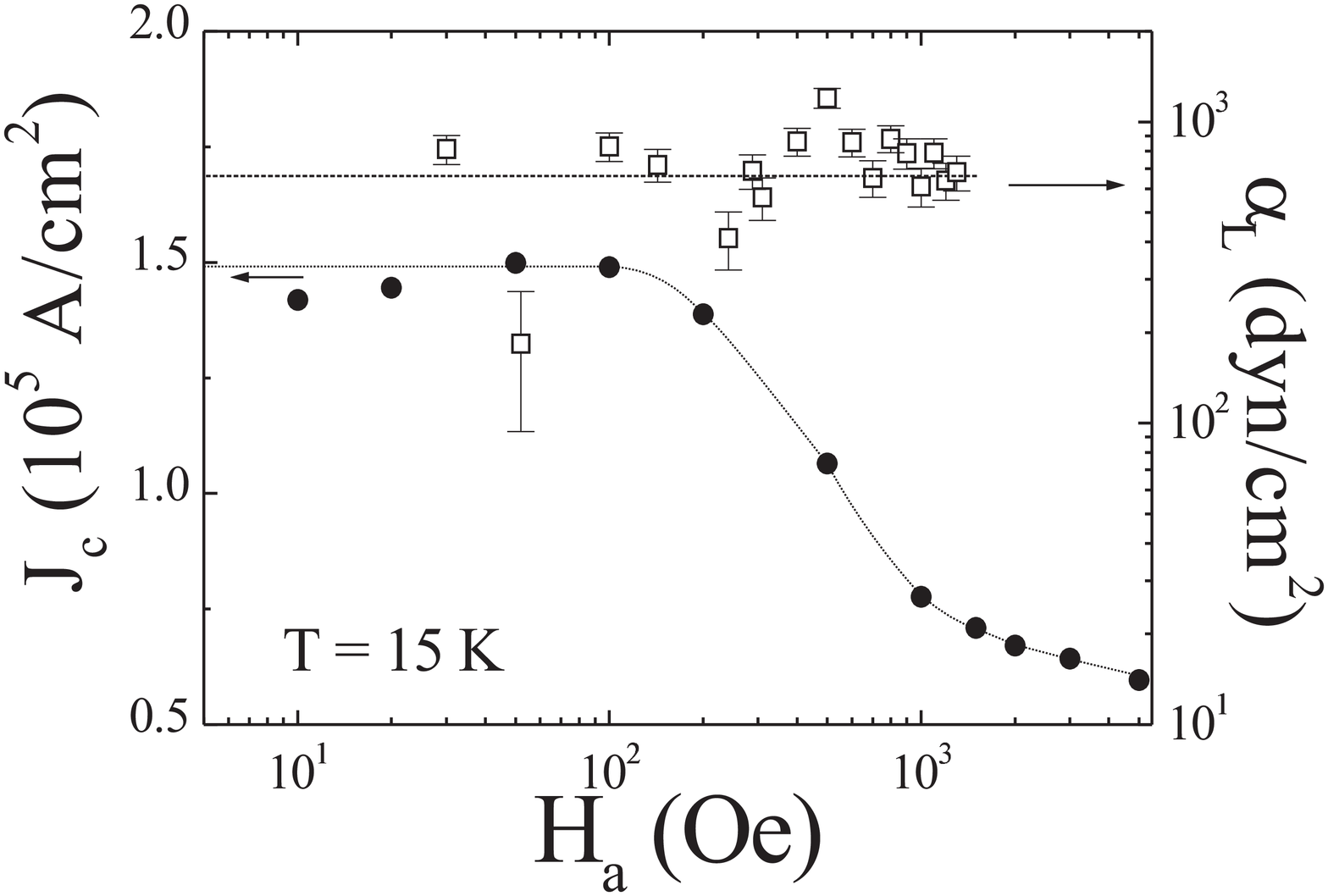}
	\caption{J$_c$ from magnetization loops and Labush coefficent $\alpha_L$ 
	from equation (5) as a funtion of the applied magnetic field. The lines are: a guide to the 
	eye for J$_c$ data and the $\alpha_L$ value from the fit in Fig. 1}
 	\label{g2}
\end{figure}
 
The current density J$_c$, obtained from (4) is also plotted in Fig. 2. It is 
interesting to point out that the J$_c$ field independent region 
(individual pinning regime) is limited to fields one order of magnitude smaller 
than those where $\alpha_L$ is seen to remain constant. As mentioned before
J$_c \propto \alpha_L$ (H$_a$,T), when J$_c$ is the true critical current. Creep 
measurements \cite{Correa} in the critical state made in the time scale of the same 
order of that used in the magnetization measurements show that the decrease
of J$_c$ as seen in Fig. 2 can be taken into account by creep effects.
Equivalent measurements in the FC structure \cite{Correa} show undetectable creep 
in agreement with the results of decoration experiment \cite{Deco} and with 
the observed constant $\alpha_L$(H$_a$,T) plotted in Fig. 2. 

Fig. 3 depicts the field dependence of $\lambda_{ac}^{2}$ for other three higher 
temperatures. The frequency independence of the results and  the low 
dissipation ($\mu'' < 0.04$) verifies that the Campbell limit is obeyed 
up to 750 Oe where dissipation is detected to increase rapidly with field 
(shadowed region in the figure). The slope of $\lambda_{ac}^{2}$ (H$_a$,T) for the 
three temperatures is seen to be independent of field (indicated by the dotted line 
with slope 1) for H$_a < 200$ Oe. It is interesting to remark the relative decrease
of $\lambda_{ac}^{2}$(H$_a$,T) with field as compared to that of the 
individual pinning limit (followed by an increase at higher fields 
\cite{Goffman}, shadowed region in the figure). 
It is surprising that the enhancement of the effective
pinning potential (better shielding) with H$_a$ takes place within the individual 
Campbell pinning limit. This is clearly seen in Fig. 4 where we have plotted
$\alpha_L$(H$_a$,T) from the data in Fig. 3, for two temperatures. 
The increase of the pinning potential with field 
strongly suggests an anomalous  softening of the elastic constant 
c$_{44}$ = (H$_a$,T) at low fields and temperatures. 

\begin{figure}[]
 	\includegraphics[angle=90, width=\columnwidth]{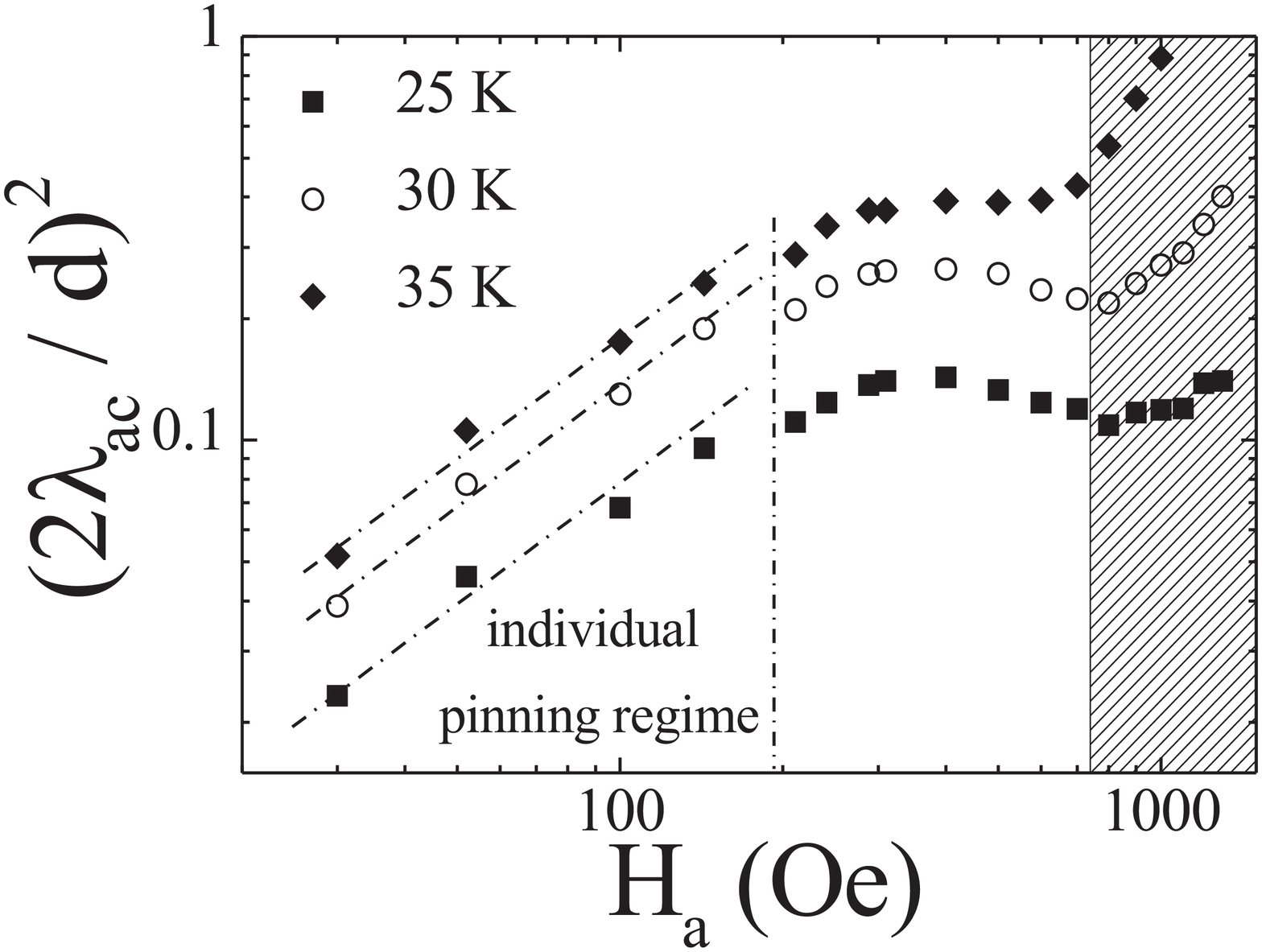}
 	\caption{Field dependence of $\lambda_{ac}^{2}$ at 25, 30 and 35K. 
	The shadowed region corresponds to a dissipative regime. The lines have 
	slope 1 indicating an individual pinning limit.}
 	\label{g3}
\end{figure}

Critical currents extracted from magnetization loops 
for T = 25K and T = 35K together with the corresponding $\alpha_L$(H$_a$,T)
from permeability measurements are shown in Fig. 4.
As observed at lower temperatures, the decrease of $J_c$
in the region of fields where $\alpha_L$ is field independent 
is also due to creep effects. Thus, the decrease in $J_c$ with H$_a$
should not be considered as an increase of the pinning correlation 
volume with field due to collective pinning effects and 
consequently should not be associated with the second peak effect.
This is further supported by a close inspection of the 
data in Fig. 4, showing that the enhancement of $\alpha_L$(H$_a$,T) 
above the individual pinning limit takes place at the field where 
the decreasing J$_c$(H$_a$) shows an inflexion point, as marked in 
the figure with an arrow. At this field a new mechanism that increases
$\alpha_L$(H$_a$,T) is switched on and, consequently, slows down the creep 
rate. This anomalous behavior of the pinning 
potential could be considered a precursor of the second peak effect.

It is interesting to point out similarities and differencies 
between the peak effect close to H$_{c2}$(T) in LTS \cite{Larkin} \cite{Kes}, 
and the second peak in Bi2212 at low fields and temperatures. In LTS, J$_c$ 
decreases with field down to a minimum at H$_{onset}$, then pinning increases
to reach a maximum at H$_{peak}$. It is shown \cite{Kes} 
that H$_{onset}$ is situated in the field region where collective pinning is 
described by a three dimensional Larkin volume. For H$_a >$ H$_{onset}$ the 
correlation volume decreases until a crossover from the collective to 
the individual pinning limit takes place at H$_{peak}$. The previous description 
implies that $\alpha_L$(H$_a$,T) decreases down to a minimum at H$_a$ = H$_{onset}$.
The enhancement of $\alpha_L$(H$_a$,T) for H$_a >$ H$_{onset}$ indicates a reduction 
of the correlation volume induced by a softening of the VS.
Agreement between theory and experiment is found \cite{Kes} only if the pinning 
correlation volume is calculated taking into account the lattice softening
induced by the dispersive nature of c$_{44}$(k). 
This is important close to H$_{c2}$ where the typical interaction length 
$\lambda_H$ = $\lambda_L /$1-b (b= H / H$_{c2}$) becomes comparable to the relevant 
elastic distortion of wave vector k of the VS. 

\begin{figure}[]
	\rotatebox{0}{\includegraphics[width=80mm]{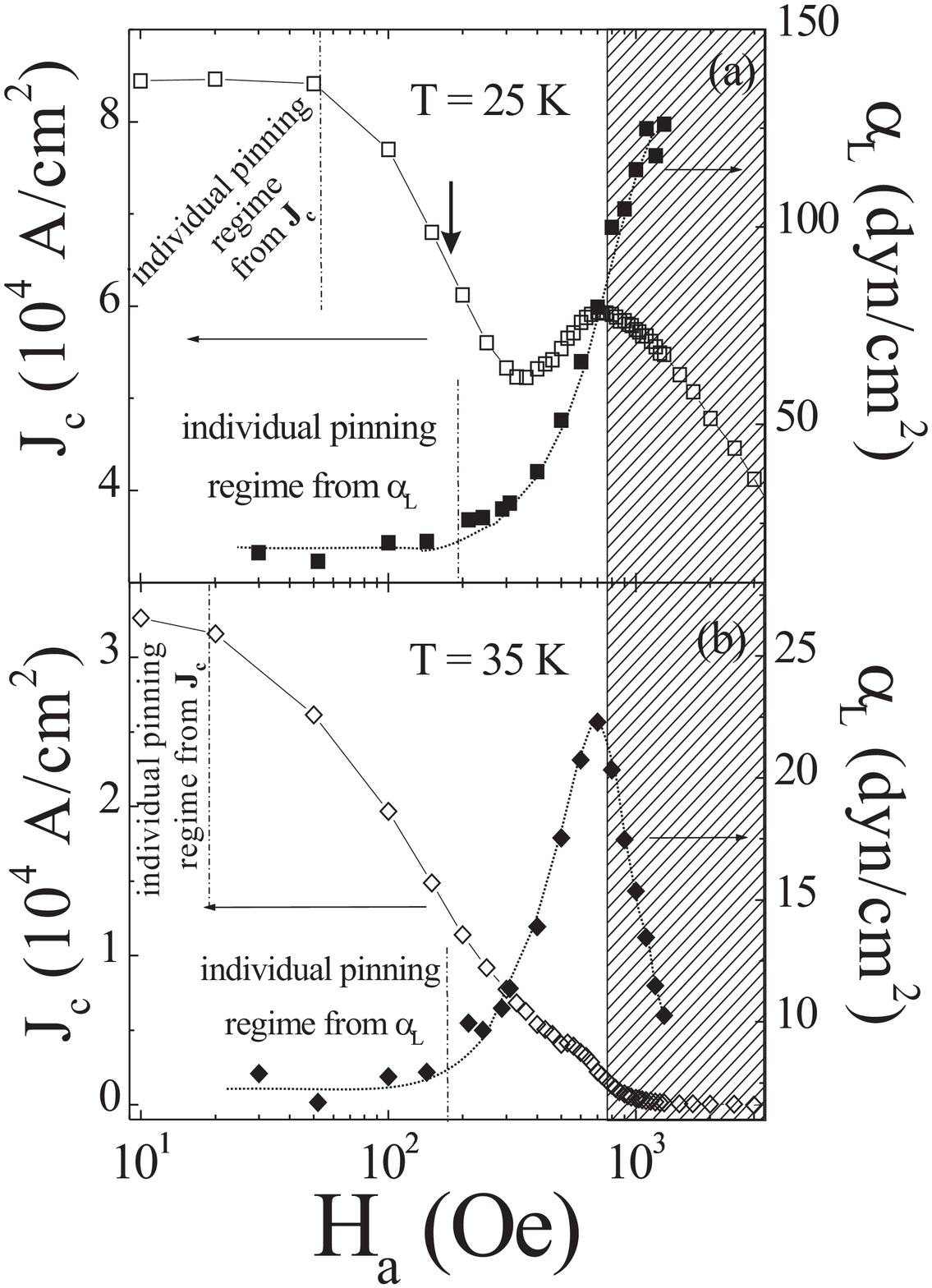}}
	\caption{J$_c$ from magnetization loops and Labush coefficent $\alpha_L$ from 
    equation (5) as a funtion of the applied magnetic field for (a) T=25K and (b) T=35K. 
    The individual pinning region (field independence) deduced for both magnitudes is indicated. 
    The arrow marks the inflection point of J$_c$(H$_a$). The shadowed region corresponds 
	to a dissipative regime.}
	\label{g4}
\end{figure}

\vspace{-1mm}

In the extreme anisotropic Bi2212, pinning is also detected to increase with field 
at the second peak. However, in this case $\alpha_L$(H$_a$,T) shows no minimum, it 
increases from a field independent value at low fields, as shown by penetration 
depth measurements. Thus the reduction of the vortex pinning correlation starts 
from the already individual Larkin volume, characterized by a one dimensional vortex 
length, L$_c$. In this limit J$_c$ is given by \cite{Larkin}

\begin{equation}
J_c \, = \, J_0(T) (\frac {\xi(T)}{L_c(T) \gamma})^2  
\end{equation}

\noindent where J$_0$(T) is the depairing current, $\xi$ is the superconducting coherence 
length, $\gamma$ is the anisotropy and L$_c$ is the Larkin correlation length in the 
field direction. 
Previous measurements of J$_c$ and $\alpha_L$(H$_a$,T) have shown \cite{Correa0D} a 
field independent temperature induced crossover from one to zero dimensional 
pinning behavior at T$_{0D} \simeq$ 20K, where L$_c$ becomes equal \cite{Correa0D} 
to the CuO interspacing distance $s$. 

In Fig. 5 the temperature dependence of $\alpha_L$(H$_a$,T) in the individual
vortex limit is depicted. The transition to the zero dimensional limit, T = T$_{0D}$, is 
evident \cite{Correa0D}. At this temperature the minimum pinning correlation 
volume (maximum $\alpha_L$(H$_a$,T) for a given temperature) is achieved. For 
T $>$ 20K the pinning is one dimensional with L$_c > s$. Thus, the increase of 
$\alpha_L$(H$_a$,T) with field, described in this paper, should be due to a 
decrease of L$_c$ induced by a softening of the elastic vortex properties. 
Following the previous discussion the maximum value that $\alpha_L$(H$_a$,T) 
can take is for 
L$_c$ = s. We have plotted in Fig. 5 the maximum values of $\alpha_L$(H$_a$,T) 
at different temperatures within the Campblell limit (maximum H$_a$ in Fig. 4 
before the shadowed area).
The linear extrapolation (dotted line) of the data below 20K
to higher temperatures strongly supports the picture discussed previously.

\vspace{-12mm}

\begin{figure}[]
	\rotatebox{270}{\includegraphics[height=63mm, width=105mm]{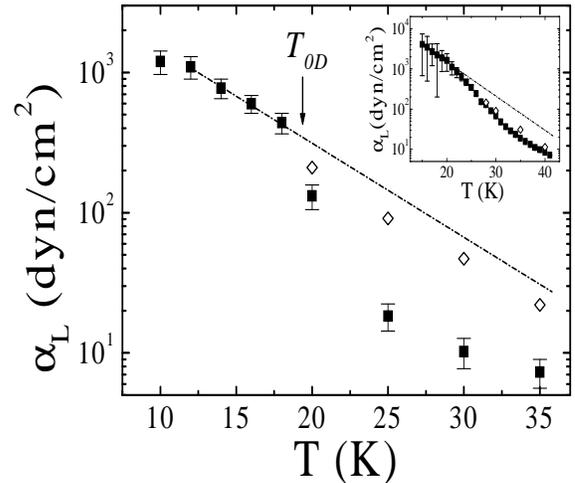}}
	\vspace{-90pt}
	\caption{Temperature dependence of $\alpha_L$(H$_a$,T) in the individual
    vortex limit. Full symbols: Field independent $\alpha_L$. Open 
    symbols: maximun value of $\alpha_L$ in the Campbell limit. The full line is an extrapolation 
    of the low temperature (below T$_{0D}$) evolution. The insert is a similar result  
    for a sample with lower deffects concentration.}
	\label{g5}
\end{figure}

It is seen from Fig. 5 that the increase of $\alpha_L$(H$_a$,T) and the
corresponding increase of J$_c$(H$_a$,T) at the second peak is due to 
of a softening of the elastic
properties of the vortices in a region of fields where its integrity
along the field is assured: individual pinning limit in the Campbell regime.
When H$_a$ is further increased J$_c$ and the effective
$\alpha_L$(H$_a$,T) are seen to decrease rapidly but the rapid increase of 
dissipation in $\mu$ indicates a crossover to a non equilibrium state.
The large dissipation is associated with currents flowing in the $c$
direction as mentioned in ref. \cite{Arribere} indicating a loss of vortex
integrity in transport properties. 

We have made measurements in two other samples with similar results: 
thermal crossover to zero dimensional pinning below T$_{0D} \simeq$ 20K, 
and a magnetic softening of the VS above T$_{0D}$.
However, it is interesting to remark that the field where dissipation
marks the end of the Campbell limit depends on each sample. In particular
we see in the insert of Fig. 5 results equivalent to those in the main 
frame. However, in this sample (that by all indications seems to be cleaner)
the dissipation appears at values of $\alpha_L$(H$_a$,T)
well below that corresponding to the zero pinning limit behavior.
This result is in agreement with a possible influence of point disorder on a 
first order transition associated with the loss of coherence in the $c$ direction.

We have shown that the increase of J$_c$ in the second peak effect is due to 
a genuine softening of the elastic properties in the individual pinning regime. 
Whether it is a manifestation of a phase transition is a subject that deserves 
more theoretical and experimental work. As mentioned in ref. \cite{Correa0D} we have 
not detected the second peak effect below 20K neither by permeability nor by J$_c$ 
measurements. Why the thermal induced transition to a zero dimensional pinning limit 
precludes the vortex softening with field remains as an open question.  

V. F. C. is fellowship holders of CONICET, Argentina.
G. N. is member of CONICET, Argentina.
Work partially supported by ANPCYT PICT97-03-00061-01116, CONICET 
PIP96/4207, Fundaci\'on Balseiro and Fundaci\'on Antorchas 1370/1-1118.

\end{document}